\documentstyle[12pt]{article}

\newcommand\tpt{\frac{t_+}{t}}
\newcommand\tmt{\frac{t_-}{t}}
\newcommand\tpm{\frac{t_+}{t_-}}
\begin{document}
\thispagestyle{empty}
\rightline{LMU-HEP-98-5}
\rightline{hep-th/9802197}
\vspace{2 truecm}
\centerline{\bf A NOTE ON INFLATIONARY STRING COSMOLOGY}

\vspace{1.2truecm}
\centerline{\bf Stefan F\" orste\footnote{
\noindent
E-mail: \tt Stefan.Foerste@physik.uni-muenchen.de\hfill}} 
\vspace{.5truecm}
\centerline{\em Sektion Physik, Universit\"at M\"unchen}
\centerline{\em Theresienstra\ss e 37, 80333 M\"unchen, Germany}

\vspace{2.2truecm}


\vspace{1truecm}

\begin{abstract}
Cosmological solutions are obtained by continuation of black D-brane
solutions into the region between the horizons. 
It is investigated whether one
can find exponential expansion when probing the cosmology with
D-branes. A unique configuration exhibiting exponential expansion
is discussed.
\end{abstract}
\vfill

\newpage

\section{Introduction}
String (or M-) theory is believed to describe physics at the Planck scale.
The only experimental window into this region is the early universe. 
Therefore it is very interesting to derive implications for the evolution 
of the universe from string theory. In [\ref{coll1}-\ref{col-last}] we give a list 
of  some references dealing with this
question. A  prominent stringy version of a cosmological model is the
pre-big-bang (PBB) scenario\cite{pbb}\footnote{Additional references
dealing with string cosmology in general and the PBB scenario in particular
can be found at Maurizio Gasperini's 
homepage: {\tt http://carmen.to.infn.it/$\tilde{\;}$gasperin/}\ .}. 
There, two solutions connected by scale
factor duality are used to describe an inflationary and a 
Friedmann-Robertson-Walker (FRW) phase of the universe. One of the
main problems in the PBB scenario is to describe the smooth phase 
transition from the inflationary to the FRW phase. This is known as the
graceful exit problem\cite{grace}. A related problem is that typically 
cosmological solutions
of low energy effective string theories run into singularities. Methods
of how to avoid those singularities are for example discussed in
\cite{lawi, kalra, bfs, maeda}. Furthermore, there are 
recent indications that the PBB scenario also 
requires an exponentially large world radius
at its onset\cite{linde}.

In this note we are going to investigate whether we can find solutions
exhibiting an inflationary phase with exponentially growing world radius.
We will deal with solutions obtained in the spirit of \cite{bf}, i.e.\
by continuing a black D-brane solution into the region between inner
and outer horizon.
Black D-brane solutions minimize the type II low energy effective action,
\begin{equation}
S = \int d^{10} x \sqrt{-g}\, \left\{ e^{-2\phi} \left[ R + 4\left( \partial \phi\right)^2
\right] - \frac{2}{\left(8-p\right) !} F^2\right\} . 
\end{equation} 
$F$ is a RR-form field strength. An object extended along $p$ spatial
dimensions couples electrically to a $p+1$ form gauge potential, 
and thus corresponds
to non-zero $p+2$ form field strength. Here, we will discuss magnetically 
charged objects and hence will have non-zero $8-p$ form field strength.
In \cite{schlager} it has been observed
that for an $(n,m)$ five-brane of type IIB theory the presence of RR 
charges increases the acceleration of the expansion near the 
big bang singularity.
Therefore, there is some hope that one can find exponential expansion
from D-brane backgrounds. In section two we will scan all the 
black D-brane backgrounds 
(continued to the region between the horizons)
for exponentially growing world radii near the singularity at the inner
horizon.
The scan includes T-dual solutions (or, in other words D-branes whose
transverse space has compact directions), and the possibility of probing
the universe with several kinds of D-branes. We will find
that there is exactly one option to obtain exponential inflation within
the given framework. This is a T-dualized D5-brane solution probed by
D0-branes. (After T-duality the D5-brane becomes a D4-brane and 
therefore probing with D0-branes makes sense.)
In section three we will investigate the inflationary model and find that
it expands in all spatial directions and is therefore a truly
ten dimensional solution. We comment on our results in a concluding section.
\section{Scan for inflationary string cosmology}
The strategy is as follows. We take a black $p$-brane solution and `go'
in the region between the horizons. In the region between the inner and the 
outer horizon, time becomes space-like and the radial distance becomes
time-like. Thus the solution (depending on the radial distance) will become
a time dependent cosmological solution.\footnote{The procedure of going
behind the outer horizon should be merely understood as an easy way
of obtaining cosmological solutions from known black brane solutions.
Nevertheless we will keep the terminology of inner and outer horizons
for convenience. 
The world-volume of the D-brane has Euclidean signature.} 
Typically,  the solutions obtained in this way possess
a big bang singularity when time is at the inner horizon and a big crunch
singularity when it reaches the outer horizon (or vice versa).
Our aim is to investigate whether we can find frames  with an 
exponential expansion in the vicinity of the
inner horizon. In such frames the universe would exhibit inflationary
expansion close to the big bang singularity.
Especially we would like to find solutions for which three 
spatial directions exponentially expand whereas the others contract or at 
least do not expand.

Before discussing the general black D-brane solution
we describe what we mean
by brane frame. A brane frame is defined such that the world volume action
of the corresponding D-brane starts off with the `canonical' Nambu-Goto
term,
\begin{equation}
\int d^{p+1} x \sqrt{g^i _{\mbox{\scriptsize brane}}} \equiv
\int d^{p+1} x e^{-\phi + \phi_0}\sqrt{g^i _{\mbox{\scriptsize string}}}
\end{equation}
where in our convention $e^{\phi_0}$ is the string coupling and  
the index $i$ on the metrics refers to the fact that they
are induced. 
From the above we obtain for the metric components 
\begin{equation} 
\left( g_{\mbox{\scriptsize brane}}\right)_{\mu\nu}
 = e^{\frac{ -2\phi + 2 \phi_0}{p+1}}
\left(g _{\mbox{\scriptsize string}}\right)_{\mu\nu}. \label{braneframe}
\end{equation}
When we probe space-time with a D-p-brane we will measure the 
corresponding brane-frame metric.

Now we will scan systematically all D-q-brane backgrounds for the 
possibility of inflationary phases in some p-brane frame.
For that we use the general form of the black brane solution in \cite{hosto}.
From their solution we obtain a cosmological solution by continuation
into the region between the two horizons.
We change the notation of \cite{hosto}
according to
\begin{equation}
t\rightarrow y \;\;\; ,\;\;\; r \rightarrow t \;\;\; ,\;\;\; r_\pm \rightarrow t_\pm ,
\end{equation}
and chose $t$ to be in the interval between $t_-$ and $t_+$, ($t_+ > t_-$). 
For a D-q-brane we find
for the metric (in the string frame)
\begin{eqnarray}
ds^2 &=& \frac{\left( \left(\tpt\right)^{7-q} -1\right)}
{ \sqrt{1 -\left( \tmt\right)^{7-q}}}dy^2 - 
\frac{dt^2}{\left( \left( \tpt \right)^{7-q} -1\right)\left( 1 - \left( \tmt\right)^{7-q}
\right)^{\frac{1}{2}+\frac{5-q}{7-q}}}\nonumber \\
& & + t^2 \left( 1 - \left( \tmt \right)^{7-q}\right)^{\frac{1}{2} - \frac{5-q}{7-q}}
d\Omega^2_{8-q} + \sqrt{ 1 - \left( \tmt \right)^{7-q}} dx^idx^i ,
\end{eqnarray}
where the sum over $i=1, \ldots , q$ is understood.
The dilaton is (for convenience we put $\phi_0 =0$, it can be reintroduced
by noting that constant shifts in $\phi$ are moduli of the low energy 
effective theory) 
\begin{equation}
e^{-2\phi} = \left( 1 - \left( \tmt\right) ^{7-q}\right)^\frac{3-q}{2} .
\end{equation}
For a magnetically charged q-brane one has an $8-q$ form field strength,
\begin{equation}
F = Q \epsilon_{8-q}
\end{equation}
where $\epsilon_{8-q}$ is the unit volume form of a $8-q$ 
sphere\footnote{For
$q=3$ one replaces $F$ by $F + *F$ in order to have a self-dual 
field strength\cite{hosto}.} and the
value for $Q$ is
\begin{equation}
Q= \frac{1}{2}\sqrt{\left( 7-q\right)^2\left( t_+ t_-\right)^{7-q}} .
\end{equation}
Now we consider the region near the inner horizon,
$t = t_- + \varepsilon $,
\begin{equation} 
1 - \left( \tmt\right)^{7-q} = (7-q) 
\frac{\varepsilon}{t_-} + {\cal O}\left( \varepsilon^2\right) .
\end{equation}
In this region the metric is approximated by
\begin{eqnarray}
ds^2 & \approx & \left(\left( \tpm\right) ^{7-q} - 1\right)
\sqrt{\frac{t_-}{(7-q)\varepsilon}} dy^2 - \frac{t_- ^{\frac{1}{2}+
\frac{5-q}{7-q}}d\varepsilon^2}{\left( \left(\tpm\right)^{7-q} -1\right)
\left( \left( 7-q\right) \varepsilon\right)^{\frac{1}{2} 
+\frac{5-q}{7-q}}}\nonumber \\
& & +t_- ^2 \left( \left( 7-q\right) \frac{\varepsilon}{t_-}\right)^{\frac{1}{2}
-\frac{5-q}{7-q}} d\Omega_{8-q}^2 +
\sqrt{\left( 7-q\right) \frac{\varepsilon}{t_-}} dx^i dx^i .
\end{eqnarray}
After compactifying the $x^i$ and $y$ directions (spanning the Euclidean
world-volume of the brane) on circles,
T-duality can be performed along the $x^i$ directions or along the
$y$ direction. The $g_{tt}$ component of the metric is not affected by
these T-dualities and will always behave like
\begin{equation}
g_{tt} \sim \varepsilon^{-\frac{1}{2} -\frac{5-q}{7-q}} .
\end{equation}
The other metric components show generically some power-like behavior
with respect to $\varepsilon$. In order to get exponential 
expansion (or contraction)
in a proper time frame the time component of the metric should go
like $\varepsilon^{-2}$.
To find inflation for a $p$-brane probe we need thus a 
dilaton behaving like
(the tilde on $\tilde{\phi}$ indicates that we allow for T-duality 
transformations)
\begin{equation}
e^{-2\tilde{\phi}} \sim \varepsilon^{\left(-\frac{3}{2} + \frac{5-q}{7-q}\right)
\left( p+1\right)}, \label{wish}
\end{equation}
where $p+1$ is the world-volume dimension of the probe.
Using the T-duality relation
\begin{equation}
\tilde{\phi} = \phi + \frac{1}{4}\log \frac{\tilde{g}}{g}
\end{equation}
we observe that
performing T-duality with respect to $n$ of the $x^i$ ($n=0,\ldots ,q$) and
$m$ times with respect to $y$ ($m=0,1$) will lead to the following 
expression
\begin{equation}
e^{-2\tilde{\phi}}\sim \varepsilon^{\frac{3-q}{2} + \frac{n}{2} - \frac{m}{2}} .
\label{is}
\end{equation}
Comparing (\ref{wish}) with (\ref{is}) we arrive at the condition
\begin{equation}
\frac{3-q}{2} +\frac{n}{2} -\frac{m}{2} =
\left( p+1 \right) \left( -\frac{3}{2} + \frac{5-q}{7-q}\right) .
\label{condition}
\end{equation}
Within the allowed parameter regions there is only one solution
to  (\ref{condition}), namely \footnote{Formally (\ref{condition}) is also
solved by $q=8$, $n=8$, $m=p=0$. The D8-brane is a solution
of massive type IIA supergravity\cite{berg} and not of the 
effective theories considered here.}
\begin{equation}
q=5 \;\;\; , \;\;\; p=0 \;\;\; , \;\;\; m=1\;\;\; , \;\;\; n=0.
\end{equation}
So, we have to start with the D5-brane background. By T-dualizing the $y$
direction this will turn into a D4-brane (with world-volume 
along $x^i$, $i=1,
\ldots ,5$). Probing this D4-brane background with D0-branes will result 
in measuring exponentially
growing scale factors near the singularity at the inner horizon.
(Later we will see that in order to get expansion and not contraction we
have to reverse the time direction. Then the singularity at the inner horizon
corresponds to a final singularity which will occur in the infinite future 
of a proper time.) 
It is quite surprising
that we find exactly one possibility to obtain inflation within the given
framework. Fortunately, the background and the probes are both objects
of type IIA theory.
\section{Inflationary string cosmology from the D5-brane}
In the previous section we gave general arguments that starting with
a (continued) D5-brane solution, performing 
T-duality along the $y$ direction
and moving to the D0-brane frame will give inflationary cosmology near
the singularity at $t=t_-$. 
Here, we will repeat the previous discussion for the D5-brane 
and analyze the result.
We start with the solution
 \begin{eqnarray}
ds^2 &=& \frac{\left( \left(\tpt\right)^{2} -1\right)}
{ \sqrt{1 -\left( \tmt\right)^{2}}}dy^2 - 
\frac{dt^2}{\left( \left( \tpt \right)^{2} -1\right)\left( 1 - \left( \tmt\right)^{2}
\right)^{\frac{1}{2}}}\nonumber \\
& & + t^2 \left( 1 - \left( \tmt \right)^{2}\right)^{\frac{1}{2}}
d\Omega^2_{3} + \sqrt{ 1 - \left( \tmt \right)^{2}} dx^idx^i ,
\end{eqnarray}
\begin{equation}
e^{-2\phi} = \left( 1 - \left( \tmt\right) ^{2}\right)^{-1} .
\end{equation}
The three-form  RR-field strength is 
\begin{equation}
F = Q \epsilon_{3}
\end{equation}
where $\epsilon_{3}$ is the unit volume form of a three-sphere and
\begin{equation}
Q= t_+ t_- .
\end{equation}
T-dualizing along the $y$ direction yields
\begin{eqnarray}
ds^2 &=& \frac{ \sqrt{1 -\left( \tmt\right)^{2}}}
{\left( \left(\tpt\right)^{2} -1\right)}dy^2 - 
\frac{dt^2}{\left( \left( \tpt \right)^{2} -1\right)\left( 1 - \left( \tmt\right)^{2}
\right)^{\frac{1}{2}}}\nonumber \\
& & + t^2 \left( 1 - \left( \tmt \right)^{2}\right)^{\frac{1}{2}}
d\Omega^2_{3} + \sqrt{ 1 - \left( \tmt \right)^{2}} dx^idx^i ,
\end{eqnarray}
\begin{equation}
e^{-2\phi} = \left( 1 - \left( \tmt\right) ^{2}\right)^{-\frac{3}{2}}
\left(\left(\tpt\right)^2 -1\right) , \label{t-dilaton}
\end{equation}
and a four form field strength (whose gauge field couples magnetically to the
D4-brane),
\begin{equation}
F \sim \epsilon_3\wedge dy . 
\end{equation}
Finally we move to the D0-brane frame by replacing $ds^2 \rightarrow
ds_0 ^2 = e^{-2\phi}ds^2$,
\begin{eqnarray}
ds^2 _0 &=& \frac{1}{ 1 -\left( \tmt\right)^{2}} dy^2 - 
\frac{dt^2}{\left( 1 - \left( \tmt\right)^{2}
\right)^{2}}\nonumber \\
& & + \frac{t^2\left( \left(\tpt\right)^2 -1 \right)}{ 1 - \left( \tmt \right)^{2}}
d\Omega^2_{3} + \frac{\left(\tpt\right)^2-1}{ 1 - \left( \tmt \right)^{2}} dx^idx^i .
\label{inflation}
\end{eqnarray}
In order to discuss the characteristics of  (\ref{inflation})
we transform to the proper time coordinate defined via
\begin{equation} 
d\tau = \pm \frac{dt}{1-\left( \tmt\right)^2}, \label{diff-proper}
\end{equation}
which is solved by
\begin{equation}
\pm \tau = t + \frac{t_-}{2} \log \frac{ t - t_-}{t+t_-} +\mbox{constant} .
\label{tau}
\end{equation}
Unfortunately we cannot obtain an analytic expression for $t$ as a function
of $\tau$ and therefore we will just give a qualitative discussion.
We reverse the time direction by choosing the lower sign in (\ref{tau}) and 
take the initial condition such that
\begin{equation}
t(\tau_0) = t_+ . \label{initial}
\end{equation}
Now the big bang is at $t = t_+$ where the scale factor in front of $dx^i dx^i$
and $d\Omega_3 ^2$ vanishes. Near the initial value $\tau_0$, $t$ 
depends linearly on $\tau$. As we approach $t_-$ all the scale factors 
grow exponentially with $\tau$ and diverge at $t\left(\tau=\infty\right) = t_-$. 
To illustrate this
we give the approximative metric near $t=t_-$. In that region
we can solve for $t(\tau)$ by
\begin{equation}
t(\tau) \approx t_- + ct_- e^{-\frac{2\tau}{t_-}} ,
\end{equation}
with $c$ being some integration constant depending on the
initial condition (\ref{initial}).
Then the metric becomes
\begin{equation}
ds_0 ^2 \approx -d\tau^2 + \frac{e^{\frac{2\tau}{t_-}}}{2c}\left\{ dy^2 +
\left( \left( \tpm\right)^2 -1\right)\left[ t_- ^2 d\Omega_3 ^2 + dx^idx^i\right]
\right\} .
\end{equation}
The dilaton behaves like
\begin{equation}
e^{-2\phi} \approx \left( \left( \tpm\right)^2 -1\right) \left(2c\right)^{-\frac{3}{2}}
e^{\frac{3\tau}{t_-}} ,
\end{equation}
and large $\tau$ is seen to correspond to weak coupling. However, for large
$\tau$ the curvature increases and $\alpha^\prime$ corrections will become
important. (Note that we are dealing with non-BPS states.)
To summarize, we have found a cosmological solution which after some
power-like expansion enters an inflationary phase with exponential expansion,
when probed with D0-branes.
For large proper time the curvature will blow up and $\alpha^\prime$ 
corrections will be important. In the presented solution all the nine
spatial directions become exponentially large. 
\section{Conclusions}
In this note we used the fact that a black D-brane corresponds to a
cosmological solution when continued to the region between the
inner and the outer horizon. From these cosmological solutions new 
solutions can be generated via T-duality. Then we searched this class
of cosmological solutions for a metric describing exponential expansion
when probed with D-branes. Surprisingly we found exactly one solution.
This is a D5-brane T-dualized to a D4-brane. When this background is
probed with D0-branes, the D0-brane like observer will see a universe
which enters an exponential expansion after some time. 
Since all nine space directions expand exponentially the universe is truly
ten dimensional. 

One may be tempted to use the presented solution in some
kind of a double-pre big bang scenario in order to achieve an exponentially
large world radius at the onset of the usual pre-big-bang scenario.
However, first of all there will be problems because one has to introduce
a second graceful exit from the exponential expansion to the 
inflationary expansion of the PBB scenario. This may be merely a technical
question. But even if one was able to solve for the graceful exit one
would have just reversed the problem raised in\cite{linde}. Now, we would
need an exponential contraction in order to get the non-observable
six space dimensions down to string scale. Before interpreting this result
as a no-go theorem for phenomenologically interesting string cosmology
one should note that we only considered a special class of cosmological
string vacua. Extending the scan for exponential expansion to more
general solutions one might be more lucky in finding some model
of phenomenological relevance. As a step towards that direction
one may consider for example intersecting black branes\cite{ara} as
a starting point.  

\bigskip
\bigskip
\noindent {\bf Acknowledgement}: It is a pleasure to thank Debashis Ghoshal,
Jacek Pawelczyk and Stefan Theisen
for many valuable discussions. I also thank Stefan Theisen for a critical
reading of the manuscript. 
The author is supported by GIF- the
German Israeli Foundation for Scientific Research.
The work is also supported in part by TMR program ERBFMX-CT96-0045.


\begin{thebibliography}{15}
\bibitem{aben} \label{coll1}   I.\ Antoniadis, C.\ Bachas, J.\ Ellis and
D.\ Nanopoulos,
{\it An expanding universe in string theory},  {\it Nucl.\ Phys.\ } 
{\bf B328} (1989) 117.
\bibitem{pbb} G.\ Veneziano, {\it Scale factor duality for classical and quantum
strings}, {\it Phys.\ Lett.\ }{\bf B265} (1991) 287;\\
M.\ Gasperini and G.\ Veneziano, {\it Pre-big-bang in string cosmology},
{\it Astropart.\ Phys.\ }{\bf 1} (1993) 317, {\tt hep-th/9211021}.
\bibitem{grace} R.\ Brustein and G.\ Veneziano, {\it The graceful exit problem
in string cosmology}, {\it Phys.\ Lett.\ }{\bf B329} (1994) 429,
{\tt hep-th/9403060};\\
R.\ Brustein and R.\ Madden, {\it Graceful exit and energy conditions in
string cosmology}, {\it Phys.\ Lett.\ }{\bf B410} (1997) 110, 
{\tt hep-th/9702043}, {\it A model of graceful exit in string cosmology},
{\it Phys.\ Rev.} {\bf D57} (1998) 712, {\tt hep-th/9708046}.
\bibitem{lawi} F.\ Larsen and F.\ Wilcek, {\it Resolution of cosmological
singularities}, Phys.\ Rev.\ {\bf D55} (1997) 4591, {\tt hep-th/9610252}.
\bibitem{kalra} S.\ Kalyana Rama, {\it Can string theory avoid cosmological
singularities?}, {\it Phys.\ Lett.} {\bf B408} (1997) 91, {\tt hep-th/9701154}.
\bibitem{bfs} K.\ Behrndt, S.\ F\"orste and S.\ Schwager,
{\it Instanton effects in string cosmology}, {\it Nucl.\ Phys.\ }{\bf B508} (1997)
391, {\tt hep-th/9704013}.
\bibitem{bf} K.\ Behrndt and S.\ F\"orste, {\it Cosmological string solutions
in four dimensions from 5d black holes}, {\it Phys.\ Lett.\ }{\bf B320} (1994)
253, {\tt hep-th/9308131}; {\it Cosmological string solutions by
dimensional reduction}, {\tt hep-th/9312167};
{\it String Kaluza-Klein Cosmology}, 
{\it Nucl.\ Phys.} {\bf B430} (1994) 441, {\tt hep-th/9403179}.
\bibitem{copeland} E.\ Copeland, A.\ Lahiri and D.\ Wands, {\it Low energy
effective string cosmology}, {\it Phys. Rev.} {\bf D50} (1994) 4868, 
{\tt hep-th/9406216}, {\it String cosmology with a time dependent 
antisymmetric tensor}, {\it Phys. Rev.} {\bf D51} (1995) 1569, 
{\tt hep-th/9410136};\\
R.\ Easther, K.\ Maeda and 
D.\ Wands, {\it Tree level string cosmology},
{\it Phys.\ Rev.} {\bf D53} (1996) 4247, {\tt hep-th/9509074};\\
E.\ Copeland, R.\ Easther and D.\ Wands, {\it Vacuum fluctuations
in axion-dilaton cosmologies}, {\it Phys.\ Rev.} {\bf D56} (1997) 874, 
{\tt hep-th/9701082};\\
E.\ Copeland, J.\ Lidsey and D.\ Wands, {\it S duality invariant
perturbations in string cosmology}, {\it Nucl.\ Phys.}
{\bf B506} (1997) 407, {\tt hep-th/9705050}.
\bibitem{levin}J.\ Levin, {\it Inflation from extra dimensions}, 
{\it Phys.\ Lett.} {\bf B343} (1995) 69, {\tt gr-qc/9411041}.
\bibitem{maeda} I.\ Antoniadis, J.\ Rizos and K.\ Tamvakis, {\it Singularity
free cosmological solutions of the effective action}, {\it Nucl.\ Phys.}
{\bf B415} (1994) 497, {\tt hep-th/9305025}, R.\ Easther and K.\ Maeda,
{\it One loop superstring cosmology and the nonsingular universe},
{\it Phys. Rev.} {\bf D54} (1996) 7252, {\tt hep-th/9605173}.
\bibitem{kaloper}N.\ Kaloper and I.\ Kogan,
{\it Cos(m)ological solutions in M theory and string theory}, 
{\tt hep-th/9711027};\\ 
N.\ Kaloper and K.\ Olive, {\it Singularities in scalar tensor cosmologies},
{\it Phys.\ Rev.} {\bf D57} (1998) 811, {\tt hep-th/9708008};\\
N.\ Kaloper, {\it Stringy Toda cosmologies}, {\it Phys.\ Rev.} {\bf D55} 
(1997) 3394, {\tt hep-th/9609087}.
\bibitem{mukherji} 
H.\ L\"u, J.\ Maharana, S.\ Mukherji and C.\ Pope,
{\it Cosmological solutions, p-branes and the
Wheeler De Witt equation}, {\it Phys.\ Rev.} {\bf D57} (1998) 2219,
{\tt hep-th/9707182};\\
H.\ L\"u, S.\ Mukherji and C.\ Pope, {\it From p-branes to cosmology},
{\tt hep-th/9612224};\\
H.\ L\"u, S.\ Mukherji, C.\ Pope and K.\ Xu,
{\it Cosmological solutions in string theories}, 
{\it Phys. Rev.} {\bf D55} (1997) 7926, {\tt hep-th/9610107}.
\bibitem{schlager} R.\ Poppe and S.\ Schwager, {\it String Kaluza-Klein
Cosmologies with RR fields}, {\it Phys.\ Lett.} {\bf B393} (1997) 51, 
{\tt hep-th/9610166}.
\bibitem{ovrut} 
A.\ Lukas, B.\ Ovrut and D.\ Waldram, {\it The cosmology of M theory
and type II superstrings}, {\tt hep-th/9802041}; {\it Stabilizing dilaton and
moduli vacua in string and M theory cosmology}, {\it Nucl.\ Phys.}
{\bf B509} (1998) 169, {\tt hep-th/9611204};
{\it String and M theory cosmological solutions with Ramond forms},
{\it Nucl. Phys.} {\bf B495} (1997) 365, {\tt hep-th/9610238};
{\it Cosmological solutions of type II string theory}, 
{\it Phys.\ Lett.} {\bf B393} (1997) 65, {\tt hep-th/9608195}.
\bibitem{saharin} A.\ Saharin, {\it Higher loop string cosmology with
moduli and antisymmetric tensor field}, {\tt hep-th/9712091}.
\bibitem{linde} M.\ Turner and E.\ Weinberg, {\it ``Pre-Big Bang
Inflation Requires Fine Tuning''}, {\it Phys.\ Rev.} {\bf D56} (1997) 4604,
{\tt hep-th/9705035};\\
A.\ Buananno {\it et al}, {\it ``Classical inhomogeneities in string cosmology''},
{\it Phys.\ Rev.} {\bf D57} (1998) 2543, {\tt hep-th/9706221};\\
N.\ Kaloper, A.\ Linde and R.\ Bousso, {\it Pre-big-bang
requires the universe to be exponentially large from the very beginning},
{\tt hep-th/9801073}.
\label{col-last}
\bibitem{hosto} G.\ Horowitz and A.\ Strominger, {\it Black strings and 
p-branes}, {\it Nucl. Phys.} {\bf B360} (1991) 197.
\bibitem{berg} E.\ Bergshoeff {\it et al}, {\it Duality of type II 7-branes and
8-branes}, {\it Nucl.\ Phys.} {\bf B470} (1996) 113, {\tt hep-th/9601150}.
\bibitem{ara} I.\ Araf\'{}eva, M.\ Ivanov, O.\ Rytchkov and I.\ Volovich,
{\it Non-extremal localised branes and vacuum solutions in M-theory},
{\tt hep-th/9802163}. 
\end{thebibliography}
\end{document}